\documentclass[aps,prl,twocolumn]{revtex4}
\usepackage{amsmath,bm,epsfig}

\def\Fbox#1{\vskip1ex\hbox to 8.5cm{\hfil\fboxsep0.3cm\fbox{%
  \parbox{8.0cm}{#1}}\hfil}\vskip1ex\noindent}  %%  {TEXT} in BOX

\newcommand{\B}[1]{{\bm{#1}}}%% Bold Roman & Greek Lower & Upper Case
\let \= \equiv \let\*\cdot \let\~\widetilde \let\-\overline

\begin{document}
\title{Statistical Physics of the Yielding Transition in Amorphous Solids}
 \author{Smarajit Karmakar, Edan Lerner and Itamar Procaccia}
\affiliation{Department of Chemical Physics, The Weizmann
 Institute of Science, Rehovot 76100, Israel }
\date{\today}

\begin{abstract}
The art of making structural, polymeric and  metallic glasses is rapidly developing with many applications. A limitation to their use is their mechanical stability:  under increasing external strain all amorphous solids respond elastically to small strains but have a finite yield stress which
cannot be exceeded without effecting a plastic response which typically leads to mechanical failure. Understanding this is crucial for assessing
the risk of failure of glassy materials under mechanical loads. Here we show that the statistics of the energy barriers $\Delta E$ that need to be surmounted changes from a probability distribution function (pdf) that goes smoothly to zero to a pdf which is finite at $\Delta E=0$. This fundamental change implies a dramatic transition in the mechanical stability properties with respect to external strain. We derive exact results for the scaling exponents that characterize the magnitudes of average energy and stress drops in plastic events as a function of system size.
\end{abstract}
\maketitle

\begin{figure}
\centering
 \includegraphics[scale = 0.54]{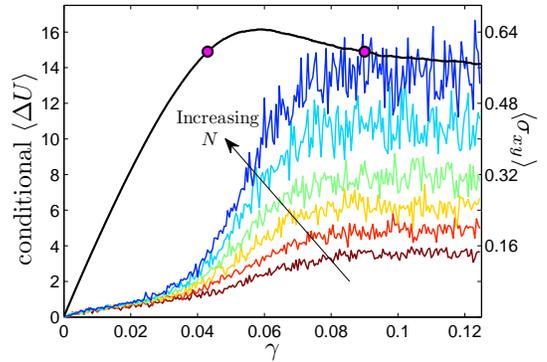}
 \caption{Color online: Evolution of the conditional mean energy drops with the loading, for all system sizes simulated, increasing from bottom to top. Superimposed (scale on the right ordinate)
 is the mean stress vs. strain curve for the largest system of $N=20164$. The magenta dots represent equi-states states but of highly different stability properties.}
 \label{conditional}
 \end{figure}

In this Letter we focus on the statistical physics of the yielding transition at very low
temperatures and quasi-static external straining conditions, (the so-called athermal quasi-static or AQS limit) where very precise
simulation results exist for the dependence of energy and stress drops in plastic events as a function of system size \cite{bunch}.
Consider Fig. \ref{conditional} which demonstrates the nature of the yielding transition. We plot here the conditional mean energy
drop in a plastic event as a function of the external strain $\gamma$ for two-dimensional systems (see below) consisting of $N$ particles,
with $N$ ranging between 484 and 20164. To read this figure properly,
one should understand that in some realizations there are no plastic events at all
at a given external stain. What is measured here is the size of the mean energy
drop {\it if} such a drop happened at an external strain value between $\gamma$ and $\gamma+d\gamma$, averaged
over numerous realizations of the random structure of the system (see below for details).
We see that in the early stages of the loading, the plastic events are localized and the amount of
energy released in events is system-size independent.
%%%%%%%%%%%%%%%%%%%%%%%%%%%%%%%%%%%%%%%%%%%%%%%%%%%%%%%
%%%%%%%%%%%%%%%%%%%%%%%%%%%%%%%%%%%%%%%%%%%%%%%%%%%%%%%%
This is followed by a smooth rise in these curves, showing an increasingly sharper transition
to the plastic flow state in which the plastic events become non-localized avalanches whose total energy release increases with the system size. This very interesting system size dependence will be quantified below. We note in passing
that the stress itself cannot be a proper order parameter;
states with the same stress level (shown for example in Fig. \ref{conditional} as two magenta circles) have very different conditional mean plastic energy drops.
Here we explore the statistical physics that is responsible for the difference between these iso-stress states,
which also have very similar potential energy and pressure.
We point out that the precise nature of this {\em strain-induced} transition from the solid-like jammed state to the steady flow state, where the plastic flow events resemble liquid-like dynamics, is still unclear.
Although the increasing availability of computational power
has recently led to many important observations and conclusions regarding the statistics of the
steady flow state \cite{bunch,09LPa,10HKLP},
a clear-cut identification of the physics
that control the \emph{approach} towards steady state, has not been presented yet. The aim of this
Letter is to close this gap and to offer some exact results. We stress that this desired analysis is best conducted
in the AQS limit since much is known there about the nature of the plastic events themselves, as these are determined
by mechanical instabilities which can be seen as a saddle node bifurcation in which the lowest eigenvalue of the Hessian
matrix going through zero \cite{04ML,99ML,10KLP}. Denote the potential energy of the system as
$U(\B r_i)$ where $\B r_i$ are the positions of the particles, and the Hessian matrix
as $H_{ij} \equiv \partial^2U /\partial \B r_i\partial \B r_j$. The Hessian is a real symmetric matrix;
we denote its lowest eigenvalue (excluding the Goldstone modes) as $\lambda_P$.
It was established \cite{04ML,10KLP} that when the external strain $\gamma$ reaches a critical value $\gamma_P$,
$\lambda_P$ vanishes with a square root singularity, i.e. such that
$\lambda_P \propto \sqrt{\gamma_P-\gamma}$.
We will show that this simple singularity determines the numerical values of a number of interesting exponents
that appear in the statistical analysis.

Below we employ a model glass-forming system with point particles of two `sizes' but of equal mass $m$
in two and three dimensions (2D and 3D respectively), interacting via a pairwise potential which had been fully
described in \cite{10KLPZ}.
The experiments performed are as follows: For un-deformed isotropic systems we measured the strain at which the first
plastic event takes place, and denoted it as $\Delta\gamma_{\rm iso}$. Each such measurement was performed on
a freshly produced amorphous solid, quenched from the high temperature liquid
at the rate of $5\times10^{-5}\frac{\varepsilon}{k_B\tau}$.
Then the AQS scheme (see \cite{10KLPZ} for details) was utilized to strain the system up to the first
mechanical instability occurring at some strain value  $\Delta\gamma_{\rm iso}$. Statistics of
 $\Delta\gamma_{\rm iso}$ were collected for a variety of system sizes, see below. In the elasto-plastic steady
 state we first strained  statistically independent systems for 100\% strain to reach stationarity,
 and then collected statistics as shown below.

\begin{figure}
 \centering
 \includegraphics[scale = 0.43]{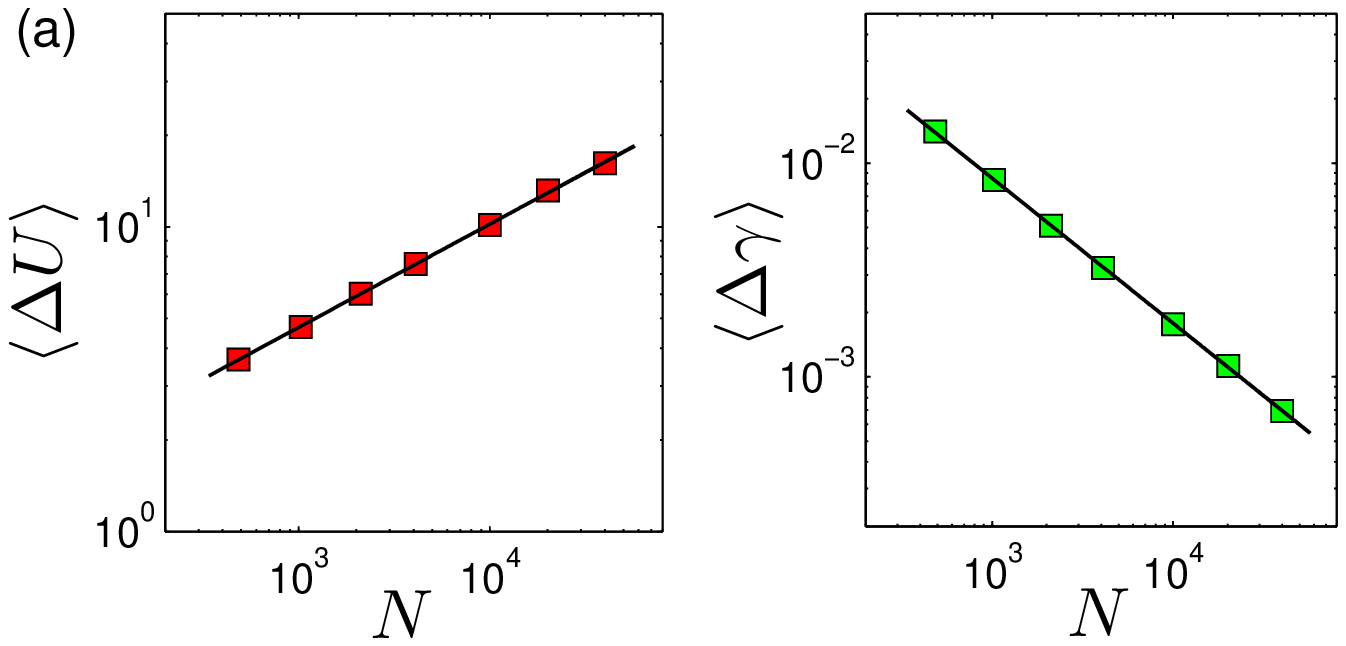}
 \includegraphics[scale = 0.43]{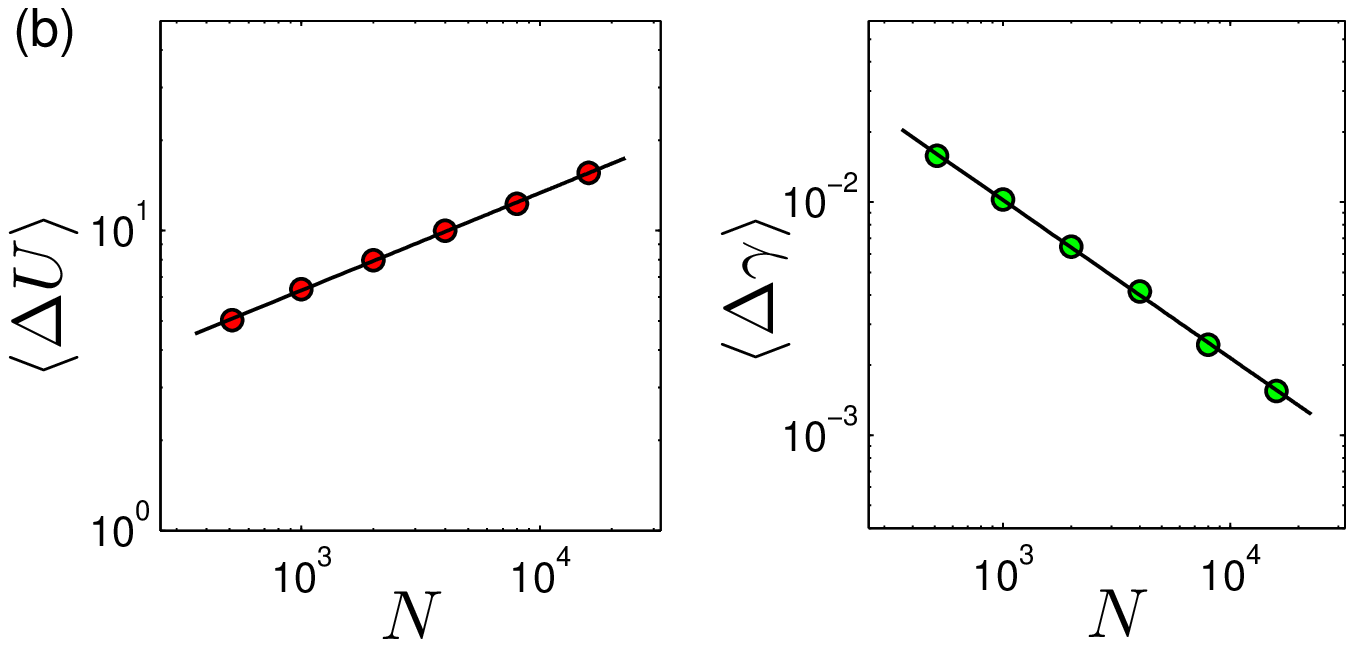}
 \caption{Color online: {panel (a).} Mean energy drop $\langle \Delta U \rangle$ and  mean strain interval
$\langle \Delta \gamma \rangle$ in 2-dimensions as functions
of system size, measured in AQS simulations of steady plastic flow of a model glass former, see text.
Panel(b): the same for three dimensions.
The continuous lines represent the scaling laws (\ref{mean1}) and (\ref{mean2}). The scaling exponents are the same in 2D and 3D.
}
 \label{meanScaling}
 \end{figure}

In the steady flow state, the statistics of the energy drops $\Delta U$, the stress drops $\Delta \sigma$ and the strain intervals between successive flow events $\Delta \gamma$ become stationary.
Quite surprisingly, one finds that the averages of these quantities obey scaling relations
with the same exponents in two and three dimensions:
\begin{eqnarray}\label{mean1}
&&\langle \Delta U \rangle \sim \bar\epsilon N^\alpha\ , \quad \langle \Delta \sigma \rangle \sim \bar s N^\beta\\
&&\langle \Delta \gamma \rangle \sim N^\beta\ . \label{mean2}
\end{eqnarray}
In Fig.~\ref{meanScaling} the mean energy drop $\langle \Delta U \rangle$
and mean strain interval $\langle \Delta \gamma \rangle$ for our model system
are displayed, together with the scaling laws (\ref{mean1}). In the upper panels we show results in two dimensions
and in the lower panel in three dimensions,
and in both $\alpha \approx 1/3$ and $\beta \approx -2/3$. A scaling relations $\alpha -\beta =1$ was already established before \cite{09LPa}. In this Letter we propose that the respective values 1/3 and -2/3 are exact.
%%%%%%%%%%%%%%%%%%%%%%%%%%%%%%%%%%%%%%%%%
%%%%%%%%%%%%%%%%%%%%%%%%%

The yielding transition is underlined by the fact that for the {\bf first} plastic event when strained from a freshly quenched isotropic state the statistics is entirely different, with $\Delta U\sim N^0$ representing a localized event without any system size dependence. On the other hand the first plastic event does not occur for any infinitesimal value of $\gamma$ and careful measurements of the mean strain interval $\langle \Delta \gamma_{\rm iso}\rangle$ that separates the un-deformed state from the first plastic event results in a scaling law
\begin{equation}
\Delta \gamma_{\rm iso}\sim N^{\beta_{\rm iso}}\ ,\quad  \beta_{\rm iso}\approx -0.62\ . \label{gameq}
\end{equation}
How can we understand the difference between $\beta$ and $\beta_{\rm iso}$ and what determines their numerical values?

\begin{figure}
 \centering
 \includegraphics[scale = 0.40]{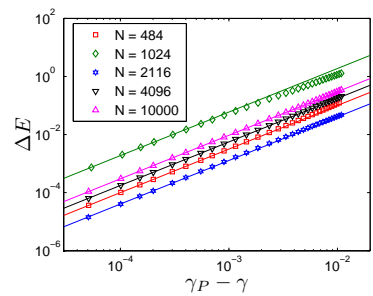}
 \caption{Color online: Scaling of energy barriers for various system sizes, see \cite{10KLPZ} for details.
 The slope of the continuous lines is 3/2.}
 \label{Evsgam}
 \end{figure}

Starting from any given mechanically stable state, at the un-deformed state or
at the steady state, the system has a set of ${\cal O}(N)$ energy barriers $\Delta E$  which are coupled to the external strain. One of those
needs to be surmounted in order to have a plastic event. In AQS conditions the one chosen will be the one
which has the smallest $\Delta \gamma_{\rm iso}$ (in the isotropic state) or the smallest $\gamma_P-\gamma$
(in the steady state). As a function of the external strain this barrier reduces until it vanishes at the saddle node
bifurcation where $\lambda_P$ vanishes \cite{04ML,99ML,10KLP}.
In Refs. \cite{06MLb,10KLPZ} it was shown that the manner in which the energy barrier vanishes is determined by the saddle node
singularity. In other words, it was established that close to $\gamma_P$
\begin{equation}
\Delta E\propto \lambda_p^3 \sim (\gamma_P-\gamma)^{3/2} \ . \label{Egam}
\end{equation}
We stress that this result is valid, sufficiently close to $\gamma_P$, equally well when starting from equilibrium,
where $\gamma_P$ represent the value of the strain for the {\em first} plastic event, or in the steady state,
where $\gamma_P$ is any value of the strain where a plastic event occurs. It turns out that the scaling law
(\ref{Egam}) is obeyed, at least in the class of models in which the potential is purely repulsive, for a very long
range of $\gamma_P - \gamma$, see Fig. \ref{Evsgam}. We will use this in this Letter, coming back to the question
of universality at the end.

In terms of distributions, the possible plastic events can occur anywhere in the system,
and the number of possible sites increases linearly with $N$. However, we realize that every time
that we observe a plastic event in AQS conditions, it is one of the {\em lowest barriers} out of ${\cal O}(N)$ barriers
that is overcome \cite{footnote}. We cannot directly measure the distribution of energy barriers but
need to concentrate on the extreme statistics regarding the minimal value. In the isotopic state, we expect the distribution
of putative strain values $x$ that could be assigned to mechanical instabilities to be Poissonian, of the form
\begin{equation}
p(x) =x^\eta h(x)/Z\ , \quad Z\equiv \int_0^\infty x^\eta h(x) \quad \eta\ge 0 \ .
\end{equation}
where $h(x)$ decays rapidly for $x\gg 1$. It is well known then that if we now take a set of $N\gg 1$ independent samples from such
a distribution, then the probability distribution $g(y,N)$ of the minimal element
of the set (denoted $y\equiv \Delta\gamma_{\rm iso}$), is the Weibull distribution \cite{39Weibull}
\begin{equation}
g(y,N) = \frac{1+\eta}{y_0}\left (\frac{y}{y_0}\right )^\eta \exp \left[-\left(\frac{y}{y_0}\right )^{1+\eta}\right] \ . \label{Weibull}
\end{equation}
In this equation $y_0\sim N^{-1/(1+\eta)}$ is the mean value of $y$ with respect to the Weibull distribution. A test of this
%%%%%%%%%%%%%%%%%%%%%%%%%%%%%%%%%%%%%%%%%%%%%%%%%%%%%%%%%%%%%%%%%
\begin{figure}
 \centering
 \includegraphics[scale = 0.38]{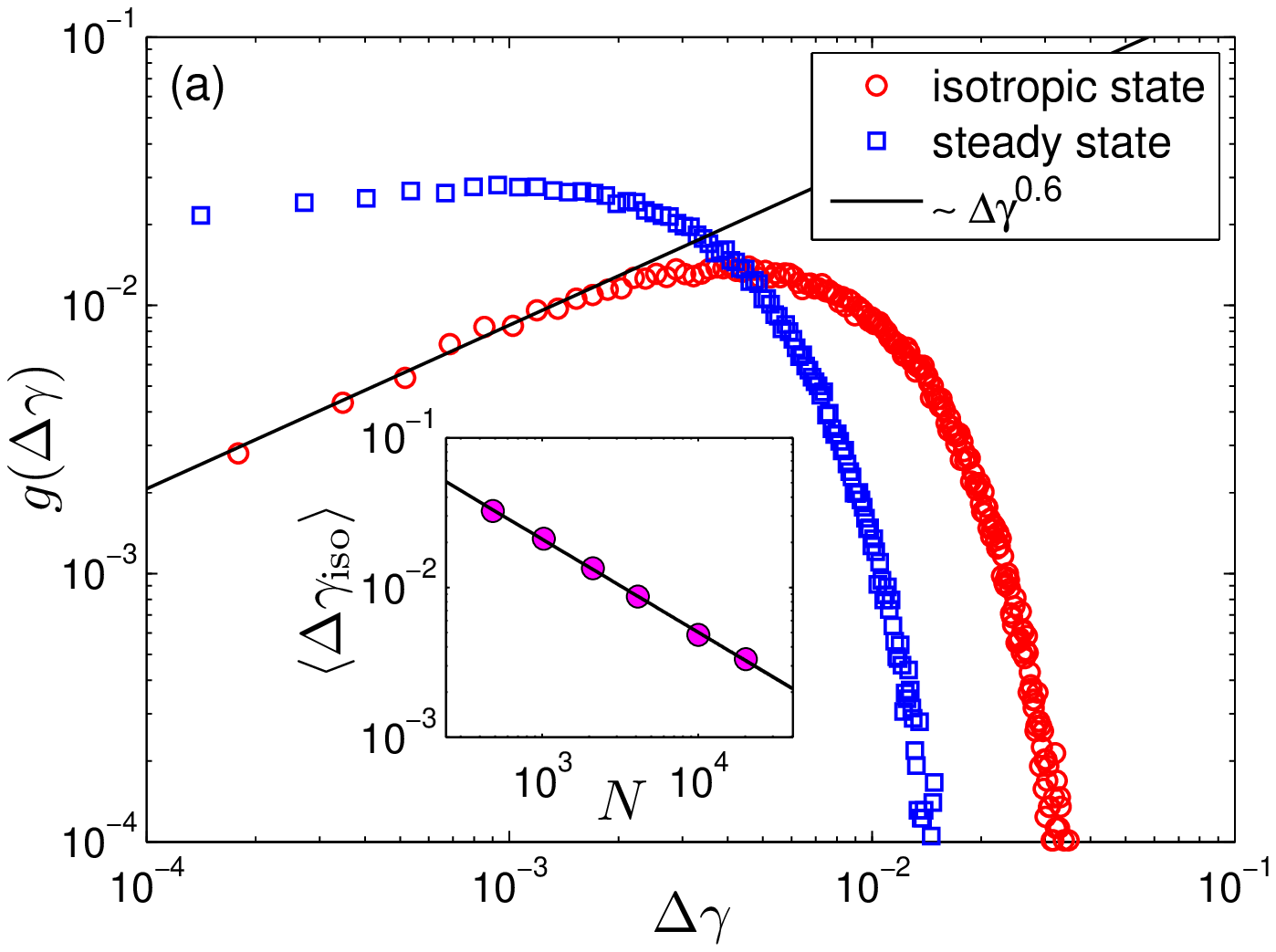}
 \includegraphics[scale = 0.38]{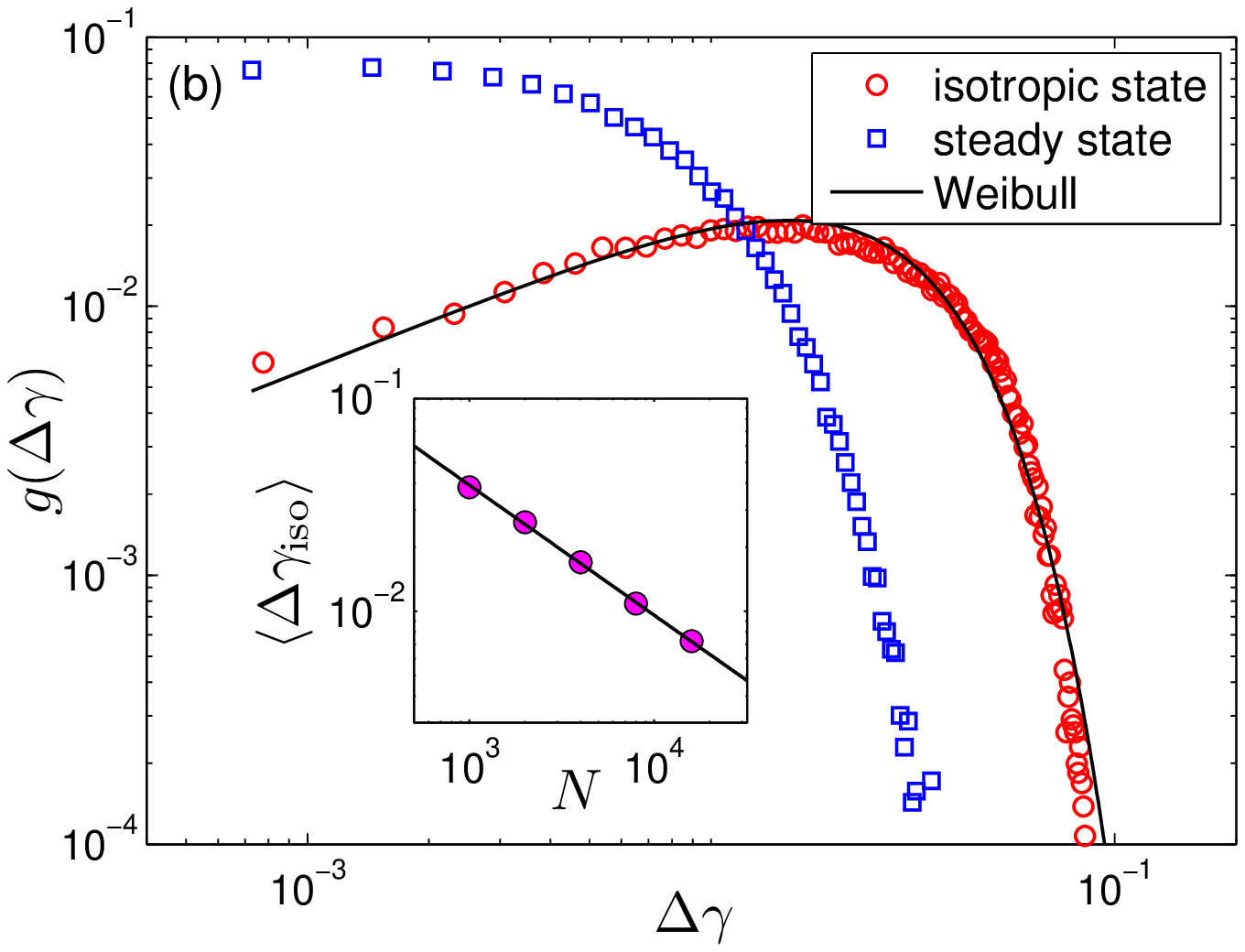}
 \caption{Color online:  the measured pdf (in 2-dimension for $N=4096$ in the panel {\bf a}) and in 3-dimensions for
$N=2000$ in panel {\bf b}) for the observed value of $x\equiv \Delta \gamma_{\rm iso}$ for isotropic systems
(in red), and of $x\equiv \Delta \gamma$ for the steady state (in blue). The scaling law for the mean shown in the inset indicates a value of $\eta\approx 0.6$ for both 2 and 3 dimensions. Note the dramatic change in the power-law tail of the distributions at small values of $\Delta \gamma$: in equilibrium the tail guarantees that the
probability to see a zero value of $\Delta E$ is zero. This is not the case in the steady state, and this is the physical
hallmark of the yielding transition. The black line through the red dots in panel {\bf b}) is the Weibull distribution (\ref{Weibull}).}
 \label{extstat}
 \end{figure}
 %%%%%%%%%%%%%%%%%%%%%%%%%%%%%%%%%%%%%%%%%%%%%%%%%%%%%
logic is presented in Fig. \ref{extstat} for systems in 3-dimensions where we show the distribution of $\Delta \gamma_{\rm iso}$ of isotropic states,
in excellent agreement with Eq.~(\ref{Weibull}) with $\eta\approx 0.6$. In 2-dimensions the scaling with $\eta\approx 0.6$ is also
recovered. We expect the distribution of
the coefficients of the scaling law Eq.~(\ref{Egam}) to be regular, implying that the probability distribution of the energy barriers has, for low values of $\Delta E$, the form
\begin{equation}
p(\Delta E)\sim (\Delta E)^{\tilde \eta}\ , \quad \tilde \eta = (2\eta-1)/3 \ .
\end{equation}
For $\eta\approx 0.6$ we find $\tilde \eta\approx 1/15>0$. This is consistent with the notion of solidity;
one expects that for a solid the probability of finding a zero barrier is strictly zero.
Finally, the mean value of the minimal values of $\Delta E$ scales
as predicted by the Weibull distribution, i.e. $\langle\Delta E\rangle\sim N^{-1/(1+\tilde \eta)}$.
Substituting this in the mean of Eq. (\ref{Egam}) we recover Eq. (\ref{gameq}) with the observed exponent
$\beta_{\rm iso}\approx 0.62$. We thus conclude that the first plastic event is dominated by extreme statistics of the minimal barriers for plasticity with a pdf of the energy barriers that goes to zero for $\Delta E\to 0$ as is expected from a solid.

The picture changes dramatically when we investigate the elasto-plastic steady state. There the distribution of strain intervals
between successive events
$\Delta \gamma$ changes qualitatively as seen in Fig. \ref{extstat}. Now the power-law tail of the
distribution is too shallow to imply that $\lim_{\Delta E\to 0} p(\Delta E) =0$. Indeed, we propose that the hallmark of the
yielding transition is that in the elasto-plastic steady state the probability to find a zero value of the energy barrier is
non-zero, cf. \cite{09RSb}. In other words, the criterion of solidity is no longer applicable. We also cannot
expect that the statistics of $\Delta\gamma$ follows the Weibull distribution, since the values of $\gamma_P$ in subsequent plastic events
become highly correlated and history dependent. Indeed the steady state distribution shown in Fig. \ref{extstat} cannot be
fitted to a Weibull distribution. On the other hand we can still expect that the numerical values
of the minimal energy barriers that need to be surmounted remain statistically independent due to the avalanches, and see next paragraph. The consequence of this is that with $N$ independent random sampling of $\Delta E$ with a finite probability
to find $\Delta E=0$, the scaling of the minimal value
must scale like $1/N$. Using this in Eq. (\ref{Egam}) leads to the proposed exact values of $\alpha=1/3$ and $\beta=-2/3$.

\begin{figure}
 \centering
 \includegraphics[scale = 0.39]{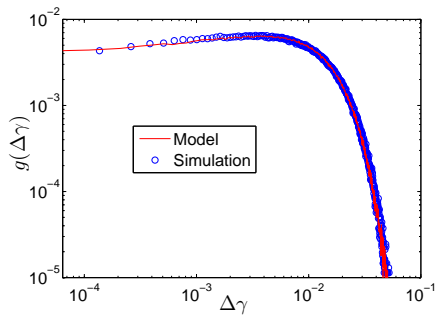}
 \caption{Color online: pdf of the observed strain intervals between avalanches compared to the converged pdf of the iterative model, see text for details. Here $q=2/3$. The agreement indicates the robustness of the model and the crucial role of the avalanches in renormalizing the pdf. Here we opted to show the pdf of strain intervals because it is the directly measurable quantity, related to the energy drop by Eq. (\ref{Egam}).}
\label{amazing}
\end{figure}

To understand the pdf of strain intervals as measured in the steady state (cf. Fig. \ref{extstat}),
we need to explain how the avalanches that occur after every plastic event refresh the statistics and renormalizes
 the pdf to the form seen in Fig. \ref{extstat}. The cumulative energy associated with an avalanche grows sub-extensively
 with the system size, (like $N^{1/3}$) and therefore the impact of these avalanches remains pertinent in the thermodynamic limit.
 To model the effect of the avalanches imagine that
 we consider a population of $N$ energy barriers $\Delta E_i$ sampled from a distribution $p(\Delta E)$ which satisfies the two conditions: ($i$) $p(\Delta E) \sim \Delta E^0$ for $\Delta E\to 0$ and ($ii$) $p(\Delta E) \to 0$ at least exponentially for $\Delta E\gg 1$. The series of plastic drops is then modeled by the following iterative steps;
at each step we repeat the following operations: 1) Find $\Delta E_{\rm min}$ and record it. 2) For every $\Delta E_i \ne \Delta E_{\rm min}$ transform,
\begin{equation}
\Delta E_i \leftarrow
\left(\Delta E_i^{2/3} - \Delta E_{\rm min}^{2/3}\right)^{3/2}.
\end{equation}
This step takes into account the fact that what is changed in the simulations is the external strain rather than
the energy barriers, and we have used Eq. (\ref{Egam}).
3) Remove $\Delta E_{\rm min}$ and re-assign a new number from $p(\Delta E)$ instead of it.
4) Randomly remove $qN$ numbers from the set ($q\in (0,1)$ is some fraction), and re-assign new numbers
instead of them from $p(\Delta E)$. This iteration scheme is readily performed numerically, leading to a converged
pdf that is shown in Fig. \ref{amazing} in comparison to the simulational pdf. We find an excellent agreement which underlines the
crucial effect of avalanches in partially destroying the correlation between subsequent values of $\Delta E_{\rm min}$. Importantly, the scaling of the mean value is invariant to the choice of $q$ in the large $N$ limit.
It should be noted that when the same iteration procedure is performed with a pdf that goes to zero at zero like $\Delta E^\eta$
(even for very small $\eta$), the resulting converged pdf is qualitatively different, preserving the scaling of the
mean value $\langle \Delta E_{\min}\rangle\sim  N^{-1/(1+\eta)}$. This stresses the importance of the fundamental
physics of the yielding transition which takes the system from a solid to liquid-like state with a qualitative change in
$\lim_{\Delta E\to 0} p(\Delta E)$.

The main point of this Letter is that the yielding transition is characterized by a qualitative change
in the nature of the pdf's of the energy barriers for a plastic event in the AQS limit. For the solid the probability to see a
vanishing energy barrier is zero. For the elasto-plastic steady state this probability is finite. Physically, this transition is
the reason for the avalanches that are observed in the steady state - there are many localized configurations with close to zero
energy barrier to surmount, and any plastic drop anywhere in the system will find it easy to cause all these configurations to
cross the instability threshold. The implication of
this qualitative change is the change in scaling exponents which are shown in Eqs. (\ref{mean2}) and (\ref{gameq}). One main
result of the Letter is the derivation of the exact values of the exponents $\alpha$ and $\beta$ in Eq. (\ref{mean2}).
We stress that this exact result rests entirely on the availability of the scaling law (\ref{Egam}). Clearly this law is
asymptotically true for $\gamma\to \gamma_P$, equally well in 2D and 3D, implying that  $\alpha$ and $\beta$ are dimension
independent. However the range of strain values over which this law pertains in the present study is quite
amazing, and there is no guarantee that this will remain true in other models which have attractive terms
in the potential of molecular degrees of freedom. It is therefore worthwhile to continue to explore the scaling
properties of different models in the AQS conditions to delineate the existence of universality classes.

This work has been supported by the Israel Science Foundation, the German Israeli Foundation,
and the Ministry of Science under the French-Israeli collaboration.


\begin{thebibliography}{99}
\bibitem{footnote}
If Eq. (7) did not have a proportionality constant it would be the lowest barrier that is alwasy
surmounted. We do not control this constant and cannot guarantee that it is the lowest barrier, rather than
one of the lowest barriers that will be surmounted.
\end{thebibliography}
\end{document}